\documentclass[preprint2,twocolumn]{aastex701}

\usepackage{mathtools}
\usepackage{todonotes}
\usepackage{xcolor}

\usepackage[T1]{fontenc}
\usepackage{amsmath,amssymb,bm,graphicx,xcolor,hyperref,orcidlink}

\newcommand{\inoisy}{\texttt{inoisy}}
\newcommand{\inoisyp}{\texttt{inoisy+}}
\newcommand{\dd}{\mathrm d}
\newcommand{\pd}{\partial}
\newcommand{\E}{\mathcal E}
\newcommand{\Rpl}{\mathcal R_{\rm pl}}

\begin{document}

\title{A Four-Dimensional Gaussian Random Field Generator for Modeling Spatiotemporal Variability in Astrophysical Sources}

\author[orcid=0000-0001-9528-1826]{Alejandro C\'ardenas-Avenda\~no}
\email[show]{cardenas@wfu.edu}
\affiliation{Department of Physics, Wake Forest University, Winston-Salem, North Carolina 27109, USA}

\correspondingauthor{Alejandro C\'ardenas-Avenda\~no}

\author[orcid=0000-0003-3984-9864]{Diego Rubiera-Garcia}
\email[]{drubiera@ucm.es}
\affiliation{Departamento de F\'isica Te\'orica and IPARCOS, Universidad Complutense de Madrid, E-28040 Madrid, Spain}

\author[orcid=0000-0002-9214-0830]{Frederic H. Vincent}
\email[]{frederic.vincent@obspm.fr}
\affiliation{LIRA, Observatoire de Paris, Université PSL, CNRS, Sorbonne Université, Université Paris Cité, CY Cergy Paris Université, 5 Place Jules Janssen, 92190 Meudon, France}

\begin{abstract}

Semi-analytic models of black-hole movies require both an emitting flow prescription and a time-dependent source variability. Existing prescriptions are often limited to either equatorial emission or time-independent sources. In this work we present a unified model for these two ingredients. First, we prescribe an off-equatorial, nongeodesic Kerr fluid rotation law by lifting an equatorial specific-angular-momentum profile to cylindrical surfaces, setting the polar component of the four-velocity to zero, normalizing the flow with the full Kerr metric at the spacetime point, and retaining the option to recover a geodesic-like plunging prescription when needed. Second, we use this velocity as the disk advection field in a four-dimensional inhomogeneous, anisotropic Mat\'ern-like Gaussian random field. We provide a parametrized model for a torus-like disk and a central jet through a single composite correlation tensor. The resulting effective model is an implementation-ready prescription for time-dependent thick-disk and disk-jet emission for relativistic studies.

\end{abstract}

\section{Introduction}
\label{sec:introduction}
\setcounter{footnote}{0}

Horizon-scale images of M87$^*$ and Sgr~A$^*$ have made black-hole imaging a direct probe of the strong-field region around compact objects \citep{EventHorizonTelescope:2019dse,EventHorizonTelescope:2022wkp}. The interpretation of these images depends not only on the spacetime geometry, but also on the geometry, motion, and variability of the emitting plasma.  This dependence is particularly important for photon-ring observables, for which the direct image is sensitive to astrophysical structure while higher-order lensed features are increasingly controlled by null geodesic dynamics \citep{Falcke:1999pj,Johnson:2019ljv,Chael:2021rjo,Cardenas-Avendano:2023dzo,Kocherlakota:2023qgo,daSilva:2023jxa,Urso:2025gos}.

The standard physical route to black-hole movies is to ray trace general relativistic magnetohydrodynamic (GRMHD) simulations~\citep{Wong:2022rqr}. This approach is dynamically self-consistent, but expensive and tied to a finite set of choices for magnetic flux state, electron thermodynamics, numerical resolution, viewing geometry, simulation duration, and radiative-transfer prescription.  A complementary route is to use semi-analytic source models~\citep{Palumbo:2022wnl, Cardenas-Avendano_2022,Chang:2024cbi}. Such models are not replacements for plasma simulations, but they make it possible to vary the mean emissivity, disk thickness, velocity field, and, when time dependence is added, stochastic correlation structure in a controlled way.

\begin{figure*}[]
\centering
\IfFileExists{Emissitivity.png}{\includegraphics[width=\textwidth]{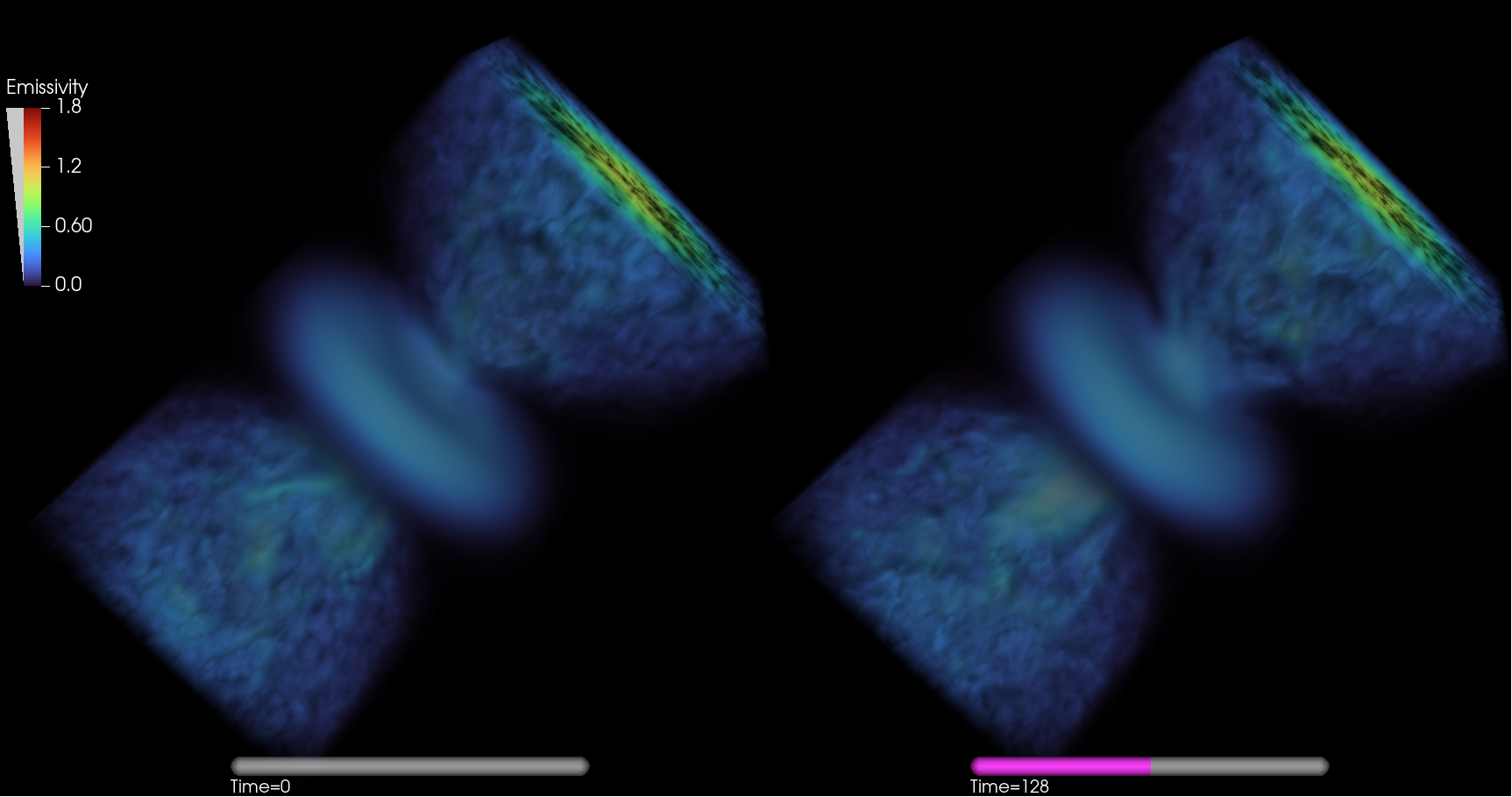}}

\caption{Two snapshots of the model emissivity using the fiducial parameters in Table~\ref{tab:model_parameters}. A single Gaussian random field with torus--jet correlations is generated by solving the four-dimensional Mat\'ern stochastic partial differential equation (Eq.~\eqref{eq:factorized_spde}), standardized to zero mean and unit variance (Eq.~\eqref{eq:standardized_field}), and mapped to positive emissivity through a lognormal transformation (Eq.~\eqref{eq:lognormal_emissivity}), with $\sigma_{\rm d}=\sigma_{\rm j}=0.2$. For this illustration, the underlying mean emissivity profiles are uniform ($\bar j_{\rm d}=\bar j_{\rm j}=1$), highlighting the stochastic variability in space and time.}
\label{fig:flow_snapshots}
\end{figure*}

The present paper combines two semi-analytic ingredients. The first is a nongeodesic kinematic prescription for a geometrically thick flow in the exterior of the Kerr spacetime. While equatorial prescriptions often interpolate between circular Keplerian-like motion and radial infall~\citep{Pu2016,THEMIS2020,Cardenas-Avendano_2022}, a geometrically thick source, however, requires a prescription away from the equatorial plane.  Without solving the relativistic Euler or magnetohydrodynamic (MHD) equations~\citep{Font2008}, and hence without specifying pressure gradients, magnetic stresses, or an external force model, there is no unique off-equatorial fluid velocity. We therefore adopt a simple stationary and axisymmetric ansatz: the angular-momentum profile is assigned on cylindrical surfaces, while the polar component of the four-velocity is set to zero. The four-velocity is then normalized with the full Kerr metric at the actual point and checked against the local timelike domain. This nongeodesic prescription can therefore be interpreted as a phenomenological force-supported fluid rotation law.

The second ingredient is a four-dimensional stochastic source model. This model is a generalization of the \inoisy{} code~\citep{Lee_2020}: time-dependent emission modeled as an inhomogeneous, anisotropic Gaussian random field whose covariance is motivated by the stochastic partial differential equation (SPDE) representation of Mat\'ern fields~\citep{Lee_2020,lindgren2011explicit}. This prescription was initially developed for equatorial (or effectively $2+1$-dimensional) sources. In contrast, for modeling thick disks, coronae, winds, and jets, a four-dimensional field is needed.

By coupling these two ingredients, we introduce \inoisyp{}\footnote{The code used for these simulations is publicly available at \url{https://github.com/alejandroc137/inoisyp} and archived on Zenodo \citep{alejandro_cardenas_avendano_2026_22833087}.} as the corresponding $3+1$-dimensional extension. The disk advection field is taken from the off-equatorial Kerr four-velocity and converted into a coordinate velocity on the source grid. As in~\cite{Lee_2020}, the stochastic part is generated by a Mat\'ern-like SPDE with a prescribed local correlation time, correlation lengths, and principal directions. The present implementation builds this correlation structure from two astrophysical building blocks: a torus-aligned disk and a collimated central jet (Fig.~\ref{fig:flow_snapshots} shows two example snapshots of the resulting stochastic simulations). These building blocks are combined at the covariance level, rather than by evolving two independent random fields by default. The resulting model is a smooth four-dimensional Mat\'ern-like random field coupled to a prescribed thick-disk and disk--jet geometry.

We work in geometric units $G=c=M=1$ and use metric signature $(-,+,+,+)$.  Greek indices refer to Kerr spacetime coordinates, while Latin indices $i,j=0,1,2,3$ refer to the Cartesian source grid used by the Gaussian random field.  Section~\ref{sec:four_velocity} constructs the off-equatorial Kerr four-velocity.  Section~\ref{sec:inoisy_plus} embeds this velocity in \inoisyp{} and defines the four-dimensional Gaussian random field.  Section~\ref{sec:discussion} summarizes the assumptions of our modeling and possible extensions.

\section{Off-equatorial Kerr four-velocity}
\label{sec:four_velocity}

We use Boyer--Lindquist coordinates $x^\mu=(t,r,\theta,\phi)$, restricted to outside the event horizon, $r>r_+=1+\sqrt{1-a^2}$, and $0<\theta<\pi$, and write
\begin{equation}
\begin{split}
\Delta&=r^2-2r+a^2,\qquad
\Sigma=r^2+a^2\cos^2\theta,\\
A&=(r^2+a^2)^2-a^2\Delta\sin^2\theta .
\end{split}
\label{eq:kerr_functions}
\end{equation}
In these coordinates, the metric components used below are
\begin{equation}
\begin{split}
g_{tt}&=-\left(1-\frac{2r}{\Sigma}\right),\qquad
 g_{t\phi}=-\frac{2ar\sin^2\theta}{\Sigma},\\
g_{rr}&=\frac{\Sigma}{\Delta},\qquad
 g_{\phi\phi}=\frac{A\sin^2\theta}{\Sigma},\\
g^{tt}&=-\frac{A}{\Sigma\Delta},\qquad
 g^{t\phi}=-\frac{2ar}{\Sigma\Delta},\\
g^{rr}&=\frac{\Delta}{\Sigma},\qquad
 g^{\phi\phi}=\frac{\Delta-a^2\sin^2\theta}
{\Sigma\Delta\sin^2\theta} .
\end{split}
\label{eq:kerr_components}
\end{equation}
While all the metric functions are evaluated at the actual spacetime point $(r,\theta)$, we will use the cylindrical radius $\rho=r\sin\theta$ as the argument of the angular-momentum profile defined below. This distinction is the essential point of the construction: the rotation law is assigned on cylindrical surfaces, while timelike normalization is imposed by the Kerr metric at the spacetime point.

This construction is deliberately not an attempt to define off-equatorial circular geodesics. In the Kerr spacetime the equatorial plane is special: a geodesic initialized with $\theta=\pi/2$ and $u^\theta=0$ remains equatorial by reflection symmetry. Here and below, $u^\theta$ denotes the Boyer--Lindquist polar component of the fluid four-velocity. Away from the equator, a geodesic with $u^\theta=0$ is generically not confined to constant $\theta$. Thus the condition $u^\theta=0$ should be interpreted as part of a fluid ansatz, supported by pressure or magnetic stresses not modeled explicitly here. The goal is to provide a closed, implementation-ready velocity field for semi-analytic emission models, while keeping the distinction between a prescribed fluid rotation law and a geodesic construction explicit.

The general equatorial four-velocity is
\begin{equation}
u^\mu\pd_\mu
=u^t\left(\pd_t-\iota\,\pd_r+\Omega\,\pd_\phi\right),
\,\,
\iota=-\frac{u^r}{u^t},
\,\,
\Omega=\frac{u^\phi}{u^t},
\label{eq:equatorial_ansatz}
\end{equation}
with $u^\theta=0$. The specific energy, angular momentum, and angular-momentum ratio are $\E=-u_t$, $L=u_\phi$, and $\ell=L/\E=-u_\phi/u_t$, respectively. For any stationary axisymmetric metric without $tr$ or $r\phi$ components, one has~\citep{1976ApJ...207..962F}
\begin{equation}
\ell(\Omega)
=-\frac{g_{t\phi}+g_{\phi\phi}\Omega}
{g_{tt}+g_{t\phi}\Omega},
\qquad
\Omega(\ell)
=\frac{g^{t\phi}-\ell g^{\phi\phi}}
{g^{tt}-\ell g^{t\phi}} .
\label{eq:ell_omega_map}
\end{equation}
These algebraic identities remain valid away from the equatorial plane whenever the denominators are nonzero.

On the equatorial plane, the Kerr circular-geodesic angular momentum is~\citep{Bardeen_1972}
$$
\ell_{\rm K}(r;a,\sigma)
=\sigma\,
\frac{r^{3/2}-2\sigma a+a^2r^{-1/2}}
{r-2+\sigma a r^{-1/2}},
\qquad
\sigma=\pm1,
$$
where $\sigma=+1$ denotes the prograde branch. We lift this expression to cylindrical surfaces and include two phenomenological parameters
\begin{equation}
\ell_{\xi,\delta}(\rho;a,\sigma)
=\xi\sigma\,
\frac{\rho^{3/2}-2\sigma a+a^2\rho^{-1/2}}
{\rho-2+\delta+\sigma a\rho^{-1/2}} .
\label{eq:ell_xi_delta}
\end{equation}

The parameter $\xi$ controls the angular-momentum normalization, while $\delta$ deforms the cylindrical profile in the strong-field region. When $\xi=1$ and $\delta=0$, Eq.~\eqref{eq:ell_xi_delta} reduces on the equatorial plane to the circular-geodesic Kerr profile. This choice cannot, however, define a timelike circular matter flow down to the horizon. The circular-geodesic four-velocity becomes null at the corresponding equatorial photon orbit and its formal continuation below that orbit is spacelike. We therefore, as in~\cite{Vincent2022}, adopt $\delta=3$ as the fiducial value. This conservative choice gives a regular, nongeodesic angular-momentum profile throughout the equatorial exterior for the Kerr spins considered here, including the full physical range $|a|<1$ when the branch label $\sigma$ is chosen consistently. For a specified spin, branch, and emitting domain, a less conservative value for $\delta$ may be chosen by requiring that the profile remain inside the local timelike domain from the outer horizon outward. On the equatorial plane the limiting obstruction is associated with the corresponding photon orbit, as the specific energy and angular momentum diverge there.

Because Eq.~\eqref{eq:ell_xi_delta} is lifted from the equatorial plane to cylindrical surfaces, equatorial regularity does not constitute a proof of timelikeness at every $(r,\theta)$. The profile must still be checked against the local Kerr light cone on the actual emitting support. The fiducial torus has small disk weight near the spin axis, where cylindrical prescriptions are most susceptible to pathologies.

The expression~\eqref{eq:ell_xi_delta} reduces to the usual equatorial expression when $\theta=\pi/2$ and to the form used in ~\cite{Vincent2022} when $a=0$. Away from the equatorial plane it is not a statement about circular geodesics at fixed $\theta$. Rather, it is a cylindrical fluid rotation law: surfaces of constant $\rho$ are assigned the same angular-momentum profile as the corresponding equatorial radius, and the resulting vector is normalized locally in the Kerr geometry. Unless explicitly stated otherwise, the symbol $\ell$ in the following equations denotes $\ell_{\xi,\delta}(\rho;a,\sigma)$ evaluated at the cylindrical radius of the field point.

The non-plunging orbiting reference flow is defined covariantly by
\begin{equation}
\hat u_\mu\dd x^\mu
=-\hat\E\left(\dd t-\ell\,\dd\phi\right),
\qquad
\hat\E=\mathcal U^{-1/2},
\label{eq:orbiting_covector}
\end{equation}
where
\begin{equation}
\mathcal U(\ell;r,\theta,a)
=-g^{tt}+2\ell g^{t\phi}-\ell^2g^{\phi\phi} .
\label{eq:U_definition}
\end{equation}
Thus, Eq.~\eqref{eq:U_definition} uses the angular momentum from Eq.~\eqref{eq:ell_xi_delta}, but the metric entering $\mathcal U$ is the full Kerr metric at $(r,\theta)$. The corresponding contravariant components are $\hat u^t=\hat\E(-g^{tt}+\ell g^{t\phi})$ and $\hat u^\phi=\hat\E(-g^{t\phi}+\ell g^{\phi\phi})$, with $\hat u^r=\hat u^\theta=0$ outside the optional plunge region.

The orbiting sector is real and timelike only where $\mathcal U>0$. The roots of $\mathcal U$ are
\begin{equation}
\begin{split}
\ell_{1,2}&=
\frac{g^{t\phi}\pm
\sqrt{(g^{t\phi})^2-g^{tt}g^{\phi\phi}}}
{g^{\phi\phi}},\\
\ell_-&=\min\{\ell_1,\ell_2\},
\qquad
\ell_+=\max\{\ell_1,\ell_2\} .
\end{split}
\label{eq:ell_roots}
\end{equation}
Outside the outer ergosurface, $r_{\rm ergo}(\theta)=1+\sqrt{1-a^2\cos^2\theta}$, the condition is $\ell_-<\ell_{\xi,\delta}<\ell_+$. Between the horizon and the outer ergosurface, the sign of $g^{\phi\phi}$ changes and the timelike domain is instead the exterior of the two roots. The cylindrical profile must therefore be checked locally; a profile that is admissible on the equatorial plane is not automatically admissible at all heights.

Since the velocity field is a fluid ansatz rather than a geodesic model, the existence of the equatorial innermost stable circular orbit (ISCO) does not by itself require the flow to plunge. A pressure-supported disk can, in principle, maintain non-geodesic rotation in a region where circular geodesics would be unstable. If one wants to keep the geodesic spirit of Cunningham's prescription~\citep{Cunningham:1975zz}, however, the parametrization also allows the inner radial branch to be treated differently. If this option is disabled, the orbiting reference flow has $\hat u^r=0$ everywhere and the model remains a purely prescribed rotation law, apart from any contribution from the separate free-fall interpolation below.

When the optional inner branch is enabled, we use the equatorial ISCO radius only as a simple spherical trigger
\[
r<r_{\rm ms},\qquad r_{\rm ms}=r_{\rm ISCO}(a,\sigma),
\]
rather than as a cylindrical condition $\rho<r_{\rm ms}$. This avoids classifying high-latitude points with small cylindrical radius as plunging material. For a point inside this sphere, define the boundary point at the same polar angle
\[
r_b=r_{\rm ms},\qquad \theta_b=\theta,\qquad
\rho_b=r_{\rm ms}\sin\theta .
\]
The frozen angular momentum is then
\[
\ell_b=\ell_{\xi,\delta}(\rho_b;a,\sigma),
\]
and the corresponding energy and angular momentum are
\[
\hat\E_b=\mathcal U(\ell_b;r_b,\theta_b,a)^{-1/2},
\qquad
\hat L_b=\hat\E_b\ell_b .
\]
These quantities define the radial reference function
\[
\Rpl=
-1-g^{tt}\hat\E_b^2
+2g^{t\phi}\hat\E_b\hat L_b
-g^{\phi\phi}\hat L_b^2 ,
\]
with all metric components in $\Rpl$ evaluated at the field point. The optional radial branch is
\begin{equation}
\hat u^r=-\sqrt{g^{rr}\Rpl}.
\label{eq:plunge_branch}
\end{equation}
This branch is retained only where the disk prescription is intended to apply, $\mathcal U(\ell_b;r_b,\theta_b,a)>0$, $\Rpl\ge0$, and the final normalized four-velocity satisfies the timelike condition below. Otherwise we set $\hat u^r=0$. This prescription should be viewed as a phenomenological inner-disk option, not as a full off-equatorial geodesic plunge.

The second reference flow is zero-angular-momentum free fall from rest at infinity. It is defined by $\bar u_t=-1$, $\bar u_\phi=0$, and $\bar u^\theta=0$, giving
\begin{equation}
\bar u^\mu=
\left(
-g^{tt},
-\sqrt{(-1-g^{tt})g^{rr}},
0,
-g^{t\phi}
\right).
\label{eq:freefall_flow}
\end{equation}
The radial square root is real outside the horizon because $(-1-g^{tt})g^{rr}=2r(r^2+a^2)/\Sigma^2>0$.

As in~\cite{Vincent2022} and~\cite{Cardenas-Avendano_2022}, the final flow is obtained by interpolating only the radial contravariant component,
\begin{equation}
u^r=\beta_r\hat u^r+(1-\beta_r)\bar u^r,
\qquad
0\le\beta_r\le1 .
\label{eq:radial_interpolation}
\end{equation}
between the optional inner reference component $\hat u^r$ and the zero-angular-momentum free-fall component $\bar u^r$. Thus, $\beta_r=1$ recovers the radial component of the orbiting reference flow, which vanishes unless the optional inner plunge is enabled, while $\beta_r=0$ gives the radial component of free fall from rest at infinity. The full zero-angular-momentum free-fall flow is recovered only if the azimuthal prescription also satisfies $\ell=0$.

For the azimuthal sector, one should not prescribe both $\ell$ and $\Omega$ independently, as 
$\Omega=\Omega(\ell_{\xi,\delta})$ (recall Eq.(\ref{eq:ell_omega_map})). Equivalently, one may adopt an angular-velocity convention and interpolate $\Omega$ directly, but then the resulting $\ell$ must be computed from Eq.~\eqref{eq:ell_omega_map} rather than set equal to Eq.~\eqref{eq:ell_xi_delta}.

The normalization $u_\mu u^\mu=-1$ fixes
\begin{equation}
u^t=
\left[
\frac{1+g_{rr}(u^r)^2}
{D(\Omega;r,\theta,a)}
\right]^{1/2},
\,\,
D=-g_{tt}-2g_{t\phi}\Omega-g_{\phi\phi}\Omega^2 .
\label{eq:ut_normalization}
\end{equation}
The off-equatorial disk four-velocity is therefore
\begin{equation}
u^\mu=\left(u^t,u^r,0,\Omega u^t\right),
\qquad
\Omega=\Omega(\ell_{\xi,\delta}) .
\label{eq:final_four_velocity}
\end{equation}
The positive root in Eq.~\eqref{eq:ut_normalization} is chosen for a future-directed flow. The final timelike condition is
\begin{equation}
D(\Omega;r,\theta,a)>0,
\qquad
\Omega_-<\Omega<\Omega_+,
\label{eq:omega_domain}
\end{equation}
where
$$
\Omega_\pm=
\frac{-g_{t\phi}\pm\sqrt{g_{t\phi}^2-g_{tt}g_{\phi\phi}}}
{g_{\phi\phi}} .
$$
If the radial sector is specified directly through $\iota=-u^r/u^t$, then
\begin{equation}
(u^t)^{-2}=D-g_{rr}\iota^2,
\qquad
0\le\iota^2<\frac{D}{g_{rr}} .
\label{eq:iota_bound}
\end{equation}
Equations~\eqref{eq:ell_xi_delta}, \eqref{eq:radial_interpolation}, and \eqref{eq:final_four_velocity} define a general four-velocity prescription. The emitting support must lie in the domain where the angular-momentum profile is finite, the reference radial flow is real, and Eq.~\eqref{eq:omega_domain} holds.

\section{A four-dimensional stochastic source}
\label{sec:inoisy_plus}

Now that we have prescribed an off-equatorial prescription for the four-velocity, we will use it to generate a fluctuating source field on a four-dimensional grid
$$
X^i=(t_{\rm s},x_{\rm s},y_{\rm s},z_{\rm s}),
\qquad
i=0,1,2,3 .
$$
The source grid is aligned with the black-hole spin axis. We identify $t_{\rm s}$ with Boyer--Lindquist time and use the Cartesian labels
$$
x_{\rm s}=r\sin\theta\cos\phi,
\qquad
y_{\rm s}=r\sin\theta\sin\phi,
\qquad
z_{\rm s}=r\cos\theta .
$$
These coordinates label the stochastic source functions; the emitter four-velocity entering radiative transfer remains the normalized Kerr vector in Eq.~\eqref{eq:final_four_velocity}. 

The off-equatorial prescription presented in the previous section was derived in the Boyer--Lindquist coordinates because the stationary and axisymmetric Killing directions are explicit and the relevant expressions take familiar forms. This choice is sufficient when the field of interest is restricted to the exterior of the black hole. However, the coordinate singularity of the Boyer--Lindquist becomes problematic for grid points at or inside the horizon. Consequently, in the numerical implementation, these points require a regularization: \(r\mapsto r_++\epsilon\), where \(\epsilon\) is a small positive parameter of order \(10^{-6}\). For applications in which the emitting flow or its correlations must extend across the horizon, the kinematic prescription developed here can instead be written in horizon-penetrating coordinates, such as ingoing Kerr--Schild coordinates.

\subsection{Four-dimensional Mat\'ern SPDE}
\label{subsec:matern_spde}

Let us define the second-order elliptic operator~\citep{Lee_2020}
\begin{equation}
\mathcal L=1-\partial_i\left[\Lambda^{ij}(X)\partial_j\right],
\qquad
\partial_i=\frac{\partial}{\partial X^i},
\label{eq:L_operator}
\end{equation}
to write an anisotropic Mat\'ern SPDE in four dimensions as
\begin{equation}
\mathcal L^\beta F(X)
=N_{\nu,4}|\Lambda(X)|^{1/4}\mathcal W(X),
\label{eq:matern_spde_general}
\end{equation}
where $\mathcal W$ is unit white noise. The tensor $\Lambda^{ij}$ fully specifies the local correlation structure of the resulting field as shown below. In four dimensions, $\beta=1+\nu/2$ and $N_{\nu,4}=4\pi\sqrt{\nu(\nu+1)}$. The parameter $\nu$ controls both the differentiability of the Mat\'ern field and the high-frequency falloff of its power spectrum. The value $\nu=1/2$ is special because it gives the exponential Mat\'ern covariance; in one-dimensional time series this is the Ornstein--Uhlenbeck, or damped-random-walk, process widely used as a baseline model for quasar and active galactic nucleus (AGN) variability~\citep{Kelly2009,MacLeod2010,Ivezic2014}. Astronomical Gaussian-process applications also use smoother Mat\'ern members with $\nu\ne1/2$, most commonly half-integer values such as $\nu=3/2,5/2,7/2,$ and $9/2$~\citep{AigrainForemanMackey2022,Jordan2021,SeikelClarkson2013,BustiClarksonSeikel2014,ElizaldeKhurshudyan2019}. These choices are popular because they remain analytically simple, admit efficient representations, and provide controlled smoothness beyond the rough damped-random-walk limit. 

Since we want to obtain a random field in $d=4$ dimensions, setting it to, e.g., $\nu=1/2$ would give $\beta=5/4$, hence a fractional power of the elliptic operator. A single second-order solve has $\beta=1$ and corresponds to the marginal $\nu=0$ case. We therefore take $\beta=2$, or $\nu=2$, and solve

\begin{equation}
\begin{split}
\mathcal LG(X)&=4\pi\sqrt6\,|\Lambda(X)|^{1/4}\mathcal W(X),\\
\mathcal LF(X)&=G(X).
\end{split}
\label{eq:factorized_spde}
\end{equation}

The two solves in Eq.~\eqref{eq:factorized_spde} are the Mat\'ern factorization. They are not a disk stage followed by a jet stage. The disk and jet enter before either solve, through the single composite tensor in Eq.~\eqref{eq:Lambda_composite} below. Our choice $\nu=2$ is therefore the practical integer-order choice appropriate to the present four-dimensional SPDE. 

If $\Lambda^{ij}$ were constant and isotropic, the high-wavenumber behavior would be $\widetilde C(k)\propto k^{-8}$. For a homogeneous but anisotropic tensor, the corresponding spectrum is instead
\[
\widetilde C(k)\propto
\left(1+k_i\Lambda^{ij}k_j\right)^{-4}.
\]
In the actual model, however, $\Lambda^{ij}$ is anisotropic, advected, spatially varying, and blended between disk and jet geometries. Thus a global Cartesian power spectrum, or an observed light-curve power spectrum after radiative transfer and projection, need not be a pure $k^{-8}$ law. The robust statement is that the present implementation defines a smoother four-dimensional integer-order Mat\'ern/SPDE model; its realized power spectrum can be measured directly from the generated field.

When $\Lambda^{ij}$ varies with position, Eq.~\eqref{eq:factorized_spde} defines an inhomogeneous Gaussian random field through the SPDE itself. The factor $|\Lambda|^{1/4}$ is retained as the local normalization inherited from the homogeneous limit, with the determinant evaluated from the full composite tensor. On a finite grid, the white noise is also scaled by the inverse square root of the four-volume of a grid cell. After generating a realization, we work with the standardized field
\begin{equation}
\widehat F(X)=
\frac{F(X)-\langle F(X)\rangle}
{\left[\langle F^2(X)\rangle-
\langle F(X)\rangle^2\right]^{1/2}} ,
\label{eq:standardized_field}
\end{equation}
where $\langle\cdot\rangle$ denotes an unweighted average over all spatial grid points and time samples in a single realization, spanning the full four-dimensional simulation domain. A convenient positive emissivity model is the mean-preserving lognormal form~\citep{Cardenas-Avendano_2022}
\begin{equation}
\begin{split}
j_{\rm s}(X)=&\,\bar j_{\rm d}(X)
\exp\left[\sigma_{\rm d}\widehat F(X)
-\frac{\sigma_{\rm d}^2}{2}\right]\\
&+\bar j_{\rm j}(X)
\exp\left[\sigma_{\rm j}\widehat F(X)
-\frac{\sigma_{\rm j}^2}{2}\right] .
\end{split}
\label{eq:lognormal_emissivity}
\end{equation}
This equation is a post-processing prescription. Since the default model has only one field $\widehat F$, the same stochastic phase modulates both deterministic channels. The separation between disk and jet emission is supplied by the envelopes $\bar j_{\rm d}$ and $\bar j_{\rm j}$ (with $\sigma_{d,j}$ their corresponding fluctuation amplitudes); if an envelope vanishes in a region, that channel does not contribute there. If both envelopes are set to unity, as in the diagnostic visualization below, Eq.~\eqref{eq:lognormal_emissivity} should be read simply as a lognormal transformation of the composite Gaussian random field rather than as a final astrophysical emissivity decomposition. Statistically independent disk and jet variability can be obtained by drawing two independent fields with $\Lambda_{\rm d}^{ij}$ and $\Lambda_{\rm j}^{ij}$ separately.

\subsection{Advection from the Kerr four-velocity}
\label{subsec:advection_velocity}

The disk advection velocity used by the Gaussian random field is the coordinate velocity $\mathbf v_{\rm d}=\dd\mathbf x_{\rm s}/\dd t_{\rm s}$ induced by Eq.~\eqref{eq:final_four_velocity}. Since $u^\theta=0$
$$
v^r_{\rm BL}=\frac{\dd r}{\dd t}=\frac{u^r}{u^t},
\qquad
\frac{\dd\phi}{\dd t}=\Omega,
$$
and therefore
\begin{equation}
\begin{split}
v_{\rm d}^x&=v^r_{\rm BL}\sin\theta\cos\phi-
\rho\Omega\sin\phi,\\
v_{\rm d}^y&=v^r_{\rm BL}\sin\theta\sin\phi+
\rho\Omega\cos\phi,\\
v_{\rm d}^z&=v^r_{\rm BL}\cos\theta .
\end{split}
\label{eq:disk_cartesian_velocity}
\end{equation}
The four-velocity and the advection velocity play different roles. The former is a normalized spacetime vector that may be used as the emitter velocity in radiative transfer. The latter is a coordinate three-velocity used to tilt stochastic correlations in the source grid. It is not normalized, and $q_0^i=(1,\mathbf v)^i$ below should not be interpreted as a four-velocity. For instance, $u^\theta=0$ does not imply $v_{\rm d}^z=0$: if the radial component is nonzero, the Boyer--Lindquist radial direction has a vertical projection away from the equatorial plane.

The jet advection velocity is phenomenological in the current implementation. The default choice is a radial outflow with optional rotation
$$
\mathbf v_{\rm j}=v_{{\rm j},p}\mathbf e_r+
\rho_{\rm s}\Omega_{\rm j}\mathbf e_\phi,
$$
with $v_{{\rm j},p}=0.5$ and $\Omega_{\rm j}=0$ in the default parameter set. This velocity controls the jet correlation pattern; it is not derived from the disk four-velocity.

For each building block $\alpha\in\{{\rm d},{\rm j}\}$, define
$$
q_{\alpha,0}^i=(1,v_\alpha^x,v_\alpha^y,v_\alpha^z),
\qquad
q_{\alpha,I}^i=(0,e_{\alpha,I}^x,e_{\alpha,I}^y,e_{\alpha,I}^z),
$$
where $I=1,2,3$ and $\{\mathbf e_{\alpha,I}\}$ is an orthonormal spatial triad. The local tensor associated with that building block is
\begin{equation}
\Lambda_\alpha^{ij}(X)=
\sum_{K=0}^3
\lambda_{\alpha,K}^2(X)q_{\alpha,K}^i(X)q_{\alpha,K}^j(X).
\label{eq:Lambda_component}
\end{equation}
The quantities $\lambda_{\alpha,K}$ are scalar correlation scales; the vectors $\mathbf e_{\alpha,I}$ specify the spatial directions with which the scales $\lambda_{\alpha,I}$ are associated.

In the default torus--jet model the code constructs $\Lambda_{\rm d}^{ij}$ and $\Lambda_{\rm j}^{ij}$ separately and then forms the single tensor
\begin{equation}
\Lambda^{ij}(X)=w_{\rm d}(X)\Lambda_{\rm d}^{ij}(X)
+w_{\rm j}(X)\Lambda_{\rm j}^{ij}(X).
\label{eq:Lambda_composite}
\end{equation}
This is the tensor that is passed to the SPDE operator in Eq.~\eqref{eq:L_operator}. The addition is therefore made at the level of the covariance tensor, not at the level of the correlation vectors. Summing vectors would introduce sign-dependent cross terms and could cancel disk and jet directions that should remain statistically distinct. Summing positive tensors preserves a positive local covariance while still evolving only one scalar stochastic field.

The weights $w_{\rm d}$ and $w_{\rm j}$ are dimensionless windows that determine where each building block contributes to the covariance. In the unnormalized option used here, $w_\alpha=W_\alpha+\epsilon_w$, so the stochastic field is preferentially structured in the torus and jet regions. The small floor $\epsilon_w$ prevents the tensor from becoming singular in low-weight regions and keeps the elliptic operator well defined on the whole grid. If normalized weights are used instead, the weights mainly interpolate the local orientation and anisotropy of the covariance.

\subsection{Torus and jet correlation geometry}
\label{subsec:disk_jet_geometry}

For the disk-like building block, we define $\rho_{\rm s}=(x_{\rm s}^2+y_{\rm s}^2)^{1/2}$ and $\varphi=\operatorname{atan}(y_{\rm s}/x_{\rm s})$, with a cylindrical basis $\{\mathbf e_\rho,\mathbf e_\phi,\mathbf e_z\}$. Instead of a cylindrical disk at each height, we use an elliptical torus cross-section centered at $R_{\rm T}$,
\begin{equation}
W_{\rm d}=\exp\left[-\frac{1}{2}
\left(\frac{\rho_{\rm s}-R_{\rm T}}{A_{\rm T}}\right)^2
-\frac{1}{2}\left(\frac{z_{\rm s}}{B_{\rm T}}\right)^2\right].
\label{eq:torus_weight}
\end{equation}
The symbols $A_{\rm T}$ and $B_{\rm T}$ are the radial and vertical widths of the torus cross-section. The unit meridional tangent and normal used in the code are
\begin{equation}
\begin{split}
\mathbf e_{\rm pol}&=\frac{1}{\mathcal N_{\rm T}}
\left[-A_{\rm T}\frac{z_{\rm s}}{B_{\rm T}}\mathbf e_\rho
+B_{\rm T}\frac{\rho_{\rm s}-R_{\rm T}}{A_{\rm T}}\mathbf e_z\right],\\
\mathbf e_{\rm n}&=\frac{1}{\mathcal N_{\rm T}}
\left[B_{\rm T}\frac{\rho_{\rm s}-R_{\rm T}}{A_{\rm T}}\mathbf e_\rho
+A_{\rm T}\frac{z_{\rm s}}{B_{\rm T}}\mathbf e_z\right],
\end{split}
\label{eq:torus_pol_normal}
\end{equation}
with
$$
\mathcal N_{\rm T}=\left[
A_{\rm T}^2\left(\frac{z_{\rm s}}{B_{\rm T}}\right)^2
+B_{\rm T}^2\left(\frac{\rho_{\rm s}-R_{\rm T}}{A_{\rm T}}\right)^2
\right]^{1/2}.
$$
At the coordinate point where $\mathcal N_{\rm T}=0$, the implementation uses the regular fallback $\mathbf e_{\rm pol}=\mathbf e_z$ and $\mathbf e_{\rm n}=\mathbf e_\rho$.

The disk correlation triad is
\begin{equation}
\begin{split}
\mathbf e_{{\rm d},1}&=\cos p\,\mathbf e_\phi+
\sin p\,\mathbf e_{\rm pol},\\
\mathbf e_{{\rm d},2}&=-\sin p\,\mathbf e_\phi+
\cos p\,\mathbf e_{\rm pol},\\
\mathbf e_{{\rm d},3}&=\mathbf e_{\rm n} .
\end{split}
\label{eq:disk_triad}
\end{equation}
Thus $\lambda_{{\rm d},1}$ controls correlations along a spiral direction on the torus surface, $\lambda_{{\rm d},2}$ along the meridional surface direction, and $\lambda_{{\rm d},3}$ across the torus surface. The quantities $\lambda_{{\rm d},I}$ are not vectors; they are the correlation lengths associated with the unit vectors $\mathbf e_{{\rm d},I}$.

For the jet building block, let $r_{\rm s}=(x_{\rm s}^2+y_{\rm s}^2+z_{\rm s}^2)^{1/2}$ and use the spherical frame $\{\mathbf e_r,\mathbf e_\vartheta,\mathbf e_\phi\}$. The default jet support is collimated around the spin axis
\begin{equation}
W_{\rm j}=\exp\left[-\frac{\rho_{\rm s}^2}{2R_{\rm j}^2(z_{\rm s})}\right],
\,\,\,
R_{\rm j}(z_{\rm s})=R_{{\rm j},0}+\alpha_{\rm j} z_{\rm{ref}} \left(\frac{|z_{\rm s}|}{z_{\rm{ref}}}\right)^k,
\label{eq:jet_weight}
\end{equation}
where $k$ is a parameter that controls the jet collimation, with $k=1$ recovering a conical profile and $k=1/2$ giving an asymptotically parabolic collimation. Intermediate values can represent observationally inferred profiles, such as $k\simeq0.58$ for M87*~\citep{2012ApJ...745L..28A}. The constant $z_{\rm ref}$ is a fixed reference length that renders the power-law argument dimensionless and keeps $\alpha_{\rm j}$ dimensionless for arbitrary collimation exponent $k$.

A helical jet triad is
\begin{equation}
\begin{split}
\mathbf e_{{\rm j},1}&=\cos\chi\,\mathbf e_r+
\sin\chi\,\mathbf e_\phi,\\
\mathbf e_{{\rm j},2}&=\mathbf e_\vartheta,\\
\mathbf e_{{\rm j},3}&=-\sin\chi\,\mathbf e_r+
\cos\chi\,\mathbf e_\phi .
\end{split}
\label{eq:jet_triad}
\end{equation}
The helical angle is regularized on the axis with
$$
\chi(\rho_{\rm s})=\chi_0\left[1-
\exp\left(-\rho_{\rm s}^2/\rho_\chi^2\right)\right],
$$
where $\chi_0$ and $\rho_\chi$ are the asymptotic helical angle and axis-regularization scale, respectively. We have now all the ingredients for this four-dimensional field. In Table~\ref{tab:model_parameters} we summarize all the parameters and provide the default values used for the examples shown below. Fig.~\ref{fig:advection_weights} shows an example of the correlation weights and the two advection fields used by the composite tensor. The top panel shows the disk advection velocity obtained from the normalized Kerr four-velocity in the equatorial slice; the bottom panel shows the phenomenological jet advection velocity in a meridional slice.

\begin{table*}[t]
\caption{Model parameters and default values for the torus--jet implementation. Lengths and times are in units of $M$. The first group controls the normalized off-equatorial Kerr fluid rotation law. The second group controls the composite tensor in Eq.~\eqref{eq:Lambda_composite}. The value $\lambda_{{\rm d},0}=-1$ is a code flag that sets the disk correlation time to the local orbital time $2\pi/|\Omega|$ (i.e., Keplerian-like). The last rows describe the diagnostic lognormal emissivity used for Fig.~\ref{fig:flow_snapshots}; for that visualization the deterministic envelopes were set to unity.}
\label{tab:model_parameters}
\centering
\setlength{\tabcolsep}{4pt}
\newcommand{\paramcol}[1]{\parbox[t]{0.20\textwidth}{#1}}
\newcommand{\defaultcol}[1]{\parbox[t]{0.18\textwidth}{#1}}
\newcommand{\meaningcol}[1]{\parbox[t]{0.54\textwidth}{#1}}
\begin{tabular*}{\textwidth}{@{}lll@{}}
\hline
\paramcol{Parameter} & \defaultcol{Default value} & \meaningcol{Meaning} \\
\hline
\paramcol{$a$} & \defaultcol{$0.94$} & \meaningcol{Dimensionless Kerr spin parameter} \\
\paramcol{$\sigma$} & \defaultcol{$+1$} & \meaningcol{Prograde angular-momentum branch} \\
\paramcol{$\xi$} & \defaultcol{$1$} & \meaningcol{Angular-momentum normalization in Eq.~\eqref{eq:ell_xi_delta}} \\
\paramcol{$\delta$} & \defaultcol{$3.0$} & \meaningcol{Cylindrical angular-momentum deformation} \\
\paramcol{$\beta_r$} & \defaultcol{$0.8$} & \meaningcol{Radial interpolation parameter in Eq.~\eqref{eq:radial_interpolation}} \\
\paramcol{$s_{\rm pl}$} & \defaultcol{no} & \meaningcol{Activates the optional spherical inner radial branch} \\
\paramcol{Timelike clamp} & \defaultcol{yes} & \meaningcol{Restricts $\Omega$ to the local timelike interval when needed} \\
\hline
\paramcol{Component} & \defaultcol{$2$} & \meaningcol{Single composite torus--jet tensor} \\
\paramcol{$\lambda_{{\rm d},A}$} & \defaultcol{$(-1,5,1.5,0.6)$} & \meaningcol{Disk time, torus-surface, meridional, and normal correlation scales} \\
\paramcol{$\lambda_{{\rm j},A}$} & \defaultcol{$(10,2.5,2.5,2.5)$} & \meaningcol{Jet time, helical, polar, and transverse correlation scales} \\
\paramcol{$R_{\rm T},A_{\rm T},B_{\rm T}$} & \defaultcol{$(12,8,5)$} & \meaningcol{Torus center radius, radial width, and vertical width} \\
\paramcol{$p$} & \defaultcol{$\pi/20$} & \meaningcol{Disk pitch angle on the torus surface} \\
\paramcol{$R_{{\rm j},0},\alpha_{\rm j}$} & \defaultcol{$(2.5,0.30)$} & \meaningcol{Jet core width and opening coefficient} \\
\paramcol{$k$} & \defaultcol{$1$} & \meaningcol{Jet collimation exponent} \\
\paramcol{$z_{\rm ref}$} & \defaultcol{$1$} & \meaningcol{Fixed reference length for the jet collimation profile} \\
\paramcol{$\chi_0,\rho_\chi$} & \defaultcol{$(\pi/20,2)$} & \meaningcol{Asymptotic helical angle and axis-regularization scale} \\
\paramcol{$v_{{\rm j},p},\Omega_{\rm j}$} & \defaultcol{$(0.5,0)$} & \meaningcol{Phenomenological jet advection parameters} \\
\paramcol{$\epsilon_w$, normalized weights} & \defaultcol{$10^{-4}$, no} & \meaningcol{Blend floor and weight-normalization choice} \\
\paramcol{$\sigma_{\rm d},\sigma_{\rm j}$} & \defaultcol{$(0.2,0.2)$} & \meaningcol{Lognormal fluctuation amplitudes} \\
\paramcol{$\bar j_{\rm d},\bar j_{\rm j}$} & \defaultcol{$(1,1)$} & \meaningcol{Disk and jet envelope, respectively}\\
\hline
\end{tabular*}
\end{table*}

\begin{figure}[]
\centering
\IfFileExists{xy_xz_weights.png}{\includegraphics[width=\columnwidth]{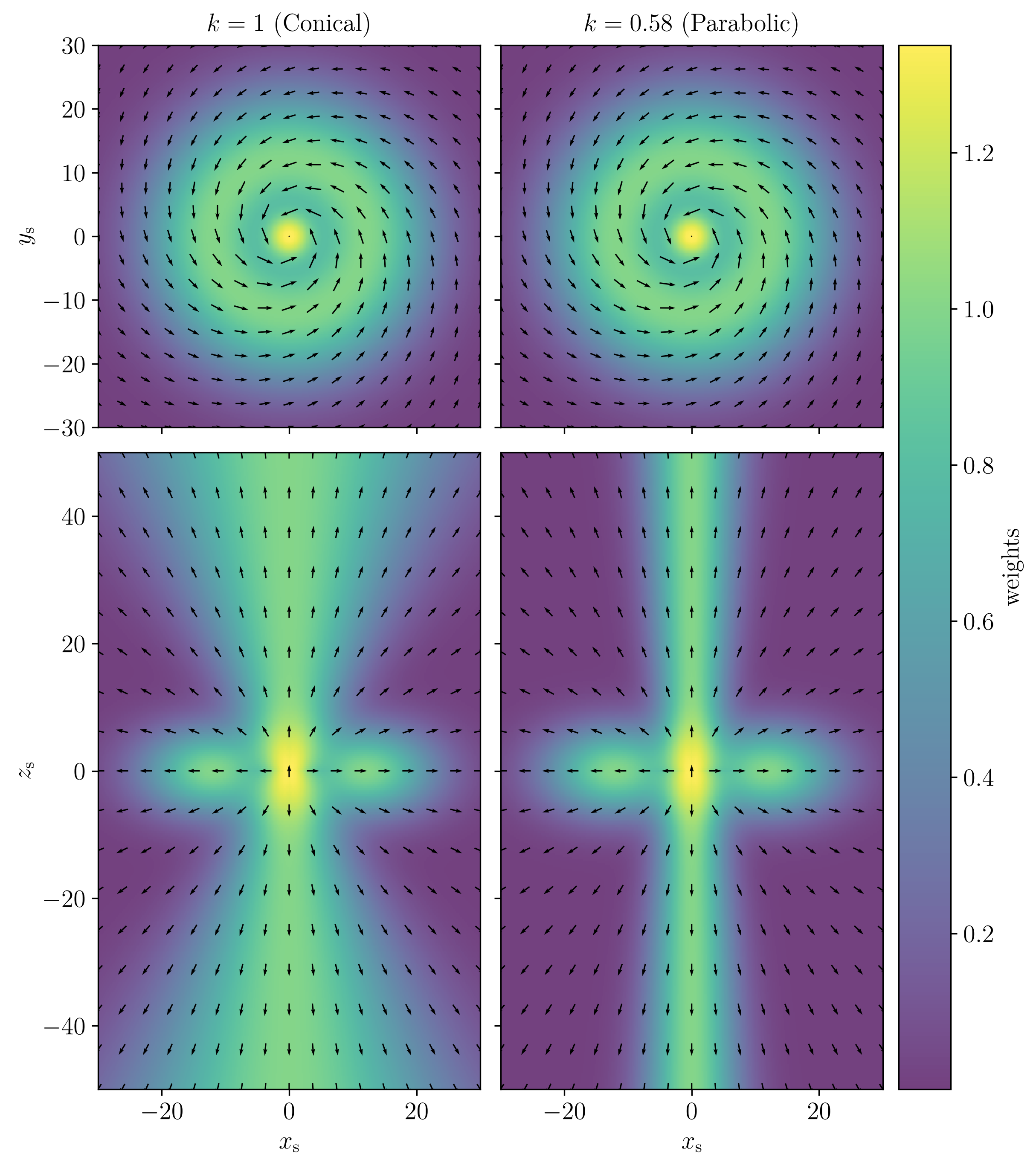}}

\caption{Torus--jet correlation geometry for conical ($k=1$, left) and parabolic ($k=0.58$, right) collimation. Both columns use $R_{{\rm j},0}=2.5M$, $\alpha_{\rm j}=0.30$, and $z_{\rm ref}=M$,
with all remaining parameters fixed to the values in Table~\ref{tab:model_parameters}. The common color scale shows the unnormalized total weight $w_{\rm d}+w_{\rm j}$. Top: equatorial
$x_{\rm s}$--$y_{\rm s}$ slices with arrows showing the disk advection
velocity $\mathbf v_{\rm d}$ induced by the normalized Kerr
four-velocity. Bottom: meridional $x_{\rm s}$--$z_{\rm s}$ slices with arrows showing the independently prescribed radial jet advection velocity $\mathbf v_{\rm j}$. All coordinates are in units of $M$.}
\label{fig:advection_weights}
\end{figure}

\begin{figure}[]
\centering
\IfFileExists{inoisyPlus_snapshots_xz.png}{\includegraphics[width=\columnwidth]{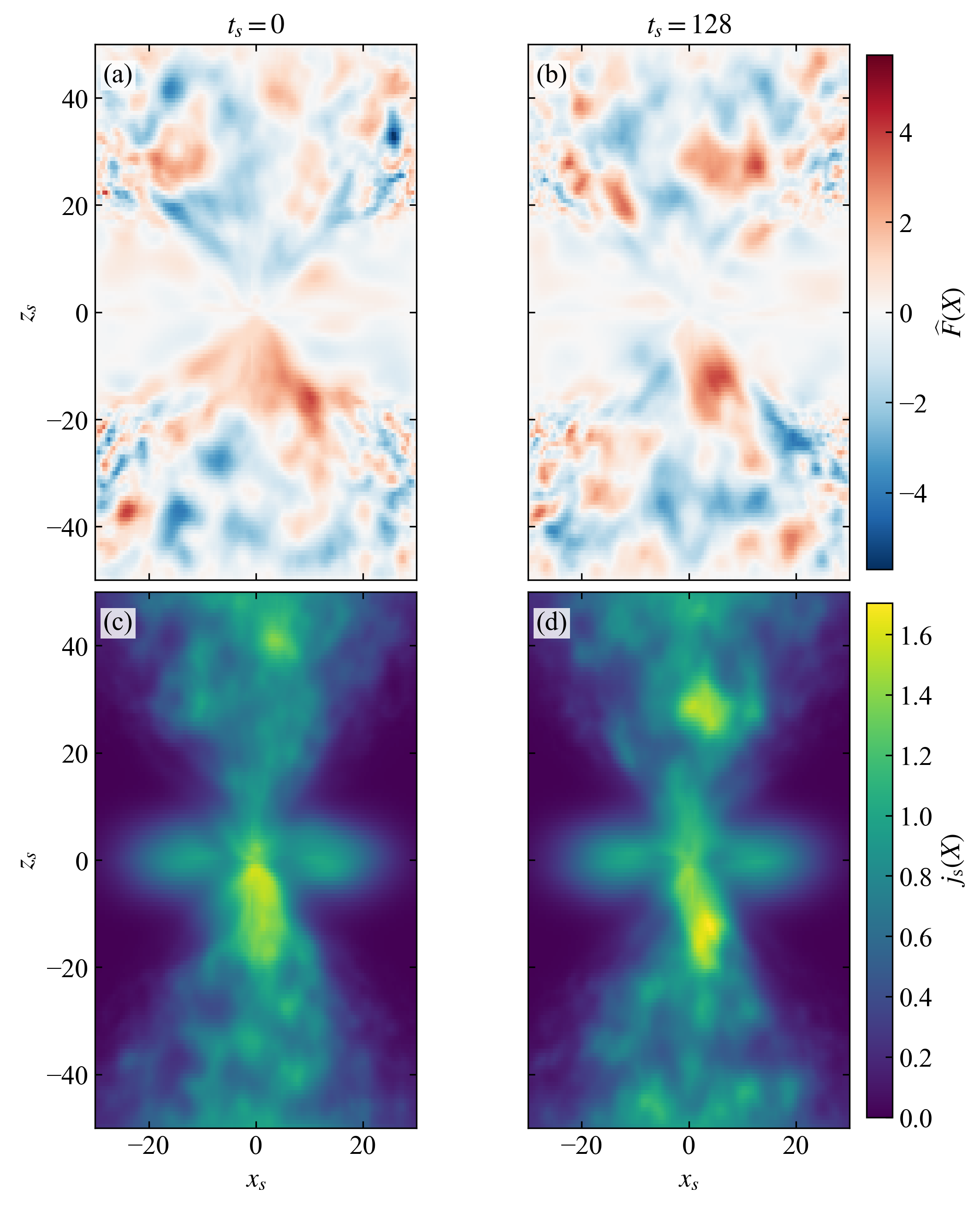}}

\caption{Meridional slices through a representative four-dimensional torus--jet realization at $t_s=0$ and $t_s=128$. Panels (a) and (b) show the standardized Gaussian random field, $\widehat F(X)$, at the two epochs, while panels (c) and (d) show the corresponding positive source field, $j_{\rm s}(X)$, obtained after applying the envelope and the lognormal mapping of Eq.~\eqref{eq:lognormal_emissivity}. The imposed envelope makes the thick equatorial torus and the collimated, two-sided jet explicit, while preserving the stochastic fluctuations of the underlying four-dimensional field. Each row uses a common color scale across the two epochs.}
\label{fig:meridionalslices}
\end{figure}

Figures~\ref{fig:flow_snapshots} and~\ref{fig:meridionalslices} show representative snapshots of the stochastic realizations after standardizing the field according to Eq.~\eqref{eq:standardized_field} and mapping it to a positive emissivity through Eq.~\eqref{eq:lognormal_emissivity}. Fig.~\ref{fig:flow_snapshots} shows a three-dimensional rendering of the resulting emissivity distribution, while Fig.~\ref{fig:meridionalslices} shows meridional slices of the same realization: the top panels show the Gaussian random field, $\widehat F(X)$, and the bottom panels show the corresponding emissivity, $j_{\rm s}(X)$. A comparison between Fig.~\ref{fig:advection_weights}'s left panel
and Fig.~\ref{fig:meridionalslices} makes the correspondence between
the analytically prescribed geometry and the resulting stochastic
source explicit (the desired thick equatorial torus
and a collimated, two-sided jet). The resulting source thus
retains the prescribed large-scale morphology while exhibiting
the fluctuations in underlying Gaussian random field. This simulation used the default parameters in Table~\ref{tab:model_parameters} on a grid of $(N_t,N_x,N_y,N_z)=(256,128,128,128)$ on $t\in[0,256)$, $x,y\in[-30,30]$, and $z\in[-50,50]$. The SPDE is solved on a grid periodic in time, while the spatial directions use the truncated finite-difference stencil at the domain boundary, i.e., homogeneous Dirichlet-type truncation. For this diagnostic image we set $\bar j_{\rm d}=\bar j_{\rm j}=1$ and $\sigma_{\rm d}=\sigma_{\rm j}=0.2$, so the displayed quantity is the lognormally transformed composite Gaussian random field rather than a radiative-transfer emissivity with separate deterministic disk and jet envelopes. 
 
To quantify the temporal variability of this torus--jet example and illustrate the spectral behavior discussed in Sec.~\ref{subsec:matern_spde}, we construct light-curve diagnostics from the standardized Gaussian field:
\begin{equation}
\begin{aligned}
L_{\rm eq}^{(W)}(t_{\rm s}) &=
\int_{\mathcal A} W(x_{\rm s},y_{\rm s},0)\,
\widehat F(t_{\rm s},x_{\rm s},y_{\rm s},0)\,
\mathrm d x_{\rm s}\,\mathrm d y_{\rm s},\\
L_V^{(W)}(t_{\rm s}) &=
\int_V W(\mathbf x_{\rm s})\,
\widehat F(t_{\rm s},\mathbf x_{\rm s})\,\mathrm d^3x_{\rm s}.
\end{aligned}
\label{eq:integrated_lightcurves}
\end{equation}
The first integral uses only the equatorial plane, $z_{\rm s}=0$;
the second includes the entire spatial grid at equal source-grid time. For each integration region, we use either $W=1$ or the additional torus--jet envelope $W=W_{\rm d}+W_{\rm j}$, with the profiles in
Eqs.~\eqref{eq:torus_weight} and \eqref{eq:jet_weight} and unit
amplitudes. Both cases retain the covariance weights used to generate
$\widehat F$. These signed integrals use coordinate area or volume
and precede the lognormal mapping of Eq.~\eqref{eq:lognormal_emissivity}
and radiative transfer.

\begin{figure*}[]
\centering
\IfFileExists{combined_psd.png}{\includegraphics[]{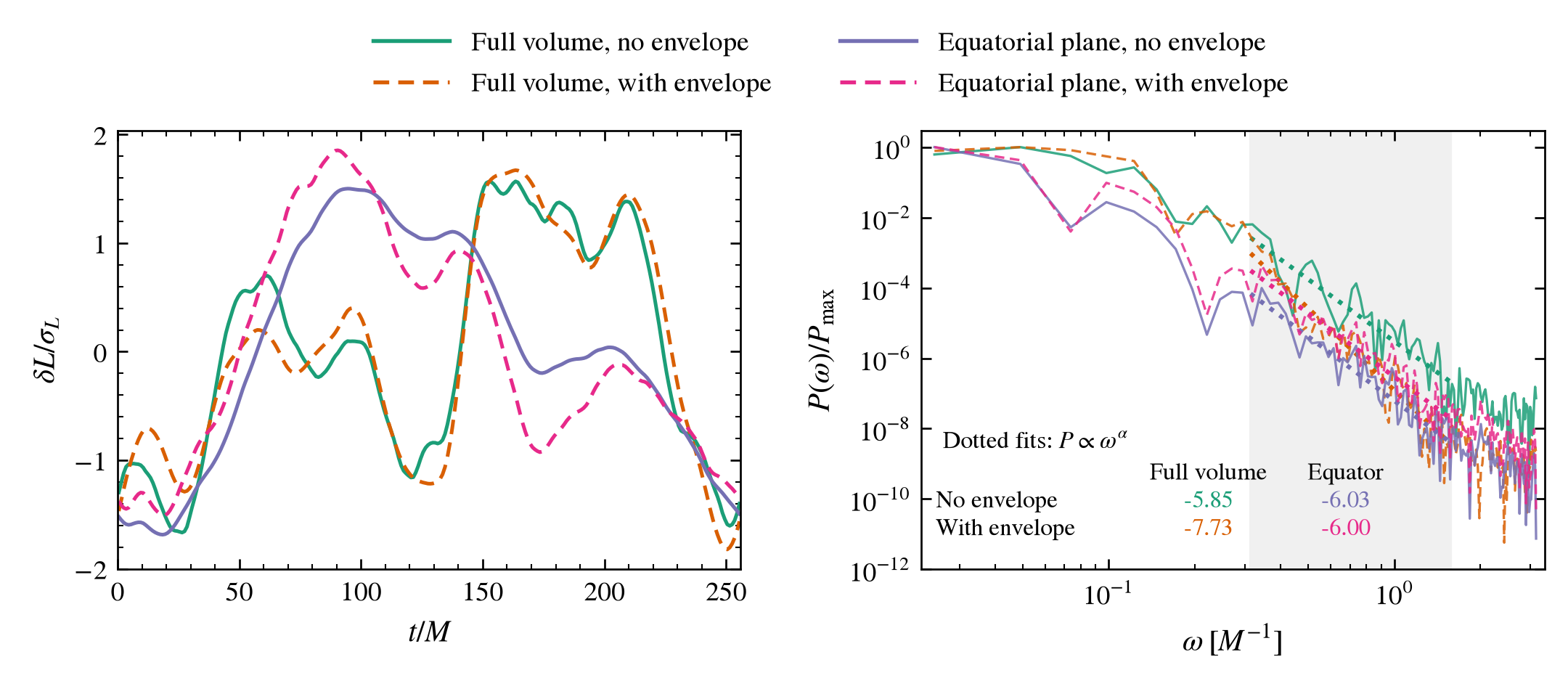}}

\caption{Spatially integrated variability of the standardized Gaussian field $\widehat F$. Green and orange curves integrate over the entire source volume while purple and magenta curves use only the equatorial plane. Dashed curves include the envelope $W=W_{\rm d}+W_{\rm j}$. Left: normalized fluctuations $\delta L/\sigma_L$, where $\delta L=L-\langle L\rangle_t$ and $\sigma_L=\langle(\delta L)^2\rangle_t^{1/2}$ are computed separately
for each curve. The fluctuations are therefore expressed in units of
each curve's own temporal standard deviation. Right: temporal power spectra estimated using a full-record Hann window and individually normalized to their maxima. Dotted lines show fits,  listed in the panel, $P(\omega)\propto\omega^\alpha$ over the shaded interval $0.314\leq\omega M\leq1.571$. All curves use the same realization, containing $256$ samples separated by $1M$.}
\label{fig:combined_psd}
\end{figure*}

Figure~\ref{fig:combined_psd} shows the resulting time series and temporal
power spectra. For each curve separately, we define
$\delta L(t_{\rm s})=L(t_{\rm s})-\langle L\rangle_t$ and
$\sigma_L=\langle(\delta L)^2\rangle_t^{1/2}$, where
$\langle\cdot\rangle_t$ denotes an average over the sampled times.
Thus $\delta L/\sigma_L$ measures departures from the temporal mean
in units of that curve's own temporal standard deviation.
Each normalized curve has zero mean and unit variance, allowing its
temporal structure to be compared independently of its absolute
amplitude. The power spectra are estimated using a full-record Hann
window and individually normalized to their maxima.

For this anisotropic, inhomogeneous realization, fits of
$P(\omega)\propto\omega^\alpha$ over $0.314\leq\omega M\leq1.571$ give $\alpha=-6.03$ and $-6.00$ for the equatorial light curves without and with the additional envelope, respectively. The corresponding volume-integrated indices are $-5.85$ and $-7.73$, demonstrating the flexibility of the model. The envelope therefore has little effect on the fitted equatorial slope, but produces a steeper volume-integrated spectrum over this interval. As discussed in Sec.~\ref{subsec:matern_spde}, the homogeneous $\omega^{-8}$ asymptote describes the full four-dimensional Fourier spectrum, but inhomogeneity, restriction to a plane, or spatial integration can define different temporal statistics. 

These simulations are computationally fast but highly memory intensive. For the example shown in Fig.~\ref{fig:flow_snapshots}, it took around $5$ hr of wall-clock time in a single node equipped with 64 AMD EPYC 9334 CPU cores with a CPU efficiency of $99\%$. However, it consumed $1.6$~TB of memory. This example shows the general trend that, although these simulations achieve excellent computational efficiency and relatively short runtimes, their primary computational limitation is memory consumption. 

Although the implementation of \inoisyp~employs the same SPDE-based approach as \texttt{inoisy}~\citep{Lee_2020} and similarly relies on the \texttt{hypre} library~\citep{hypre1,hypre2}, the generalization from an effectively $2+1$-dimensional source to a fully $3+1$-dimensional field entails substantially more than the mere introduction of an additional loop index. The original \texttt{inoisy} implementation uses the \texttt{Struct} interface on a three-dimensional structured grid and assembles a $19$-point stencil for the variable-coefficient operator. Here the source field lives on $X^i=(t_{\rm s},x_{\rm s},y_{\rm s},z_{\rm s})$, and the local covariance is a full $4\times4$ tensor. The discretized operator therefore contains four pure second-derivative directions and six mixed-derivative pairs, giving a $33$-point stencil. This required rebuilding the grid, stencil, matrix, and vector assembly using the \texttt{SStructMatrix} interface and solving the resulting object through the \texttt{ParCSR} solvers. Because the spatially varying tensor and mixed derivatives generally make the assembled operator nonsymmetric, the default solve uses \texttt{GMRES} with \texttt{BoomerAMG} preconditioning, with \texttt{PCG} retained only as an option for more symmetric test cases. Thus the main computational cost of the four-dimensional extension is not only the larger number of grid cells, but also the larger stencil, more general matrix interface, and increased memory footprint.

Convergence of both linear solves is essential to the numerical implementation of the factorized SPDE in
Eq.~\eqref{eq:factorized_spde}, since the first-stage solution supplies
the source for the second stage. For the example realization shown in
this work, each solve independently satisfied a prescribed
relative-residual tolerance of $10^{-4}$. This residual measures the
normalized mismatch in the discretized equations and therefore provides
a direct convergence criterion. The first and second stages required
$22$ and $35$ iterations, respectively, reaching final relative residuals
of $9.73\times10^{-5}$ and $8.83\times10^{-5}$. These measurements
explicitly verify that the complete two-stage calculation achieved the
requested residual tolerance at the adopted grid resolution, providing
a quantitative convergence check for the reported stochastic realization.

\section{Discussion}
\label{sec:discussion}

In this work, we have presented a unified semi-analytic prescription for thick, time-dependent black-hole source models. The first part is an off-equatorial, fluid rotation law in Kerr spacetime. The construction distinguishes the cylindrical radius used to prescribe angular momentum from the spacetime point at which the Kerr metric normalizes the four-velocity. This distinction is essential away from the equator: a profile that is valid on the equatorial plane need not lie inside the local timelike domain at arbitrary $(r,\theta)$. The ansatz $\ell=\ell(\rho)$ with $u^\theta=0$ is not a geodesic closure; it is a compact phenomenological prescription for a stationary, axisymmetric thick flow in the absence of an explicit pressure, magnetic, or force model.

The second part is the use of this velocity field in \inoisyp, a generalization of the prescription presented in~\cite{Lee_2020}. The Kerr four-velocity supplies the emitter motion and, through Eq.~\eqref{eq:disk_cartesian_velocity}, the advection vector entering the four-dimensional Gaussian random field. This gives a direct link between the mean flow model and the stochastic variability model. The extension is then achieved by promoting the source from an effectively equatorial field to a four-dimensional field and by assigning the field a composite torus--jet correlation tensor. The disk and jet are therefore two geometrical building blocks of the covariance, not two sequential SPDE stages.

This stochastic prescription is deliberately phenomenological. It does not solve the relativistic Euler equations, impose conservation laws on the random field, or determine the disk and jet velocity fields dynamically. Likewise, the optional inner radial branch is a modeling choice for applications that want an ISCO-motivated inflow region; disabling it leaves a purely force-supported rotation law. The purpose of the stochastic prescription is instead to provide a controlled, computationally efficient model for source variability. The choice $\nu=2$ in four dimensions is both a modeling and practical choice: it avoids fractional operators and the singular $\nu=0$ limit, while reducing the problem to repeated second-order elliptic solves that can be handled by standard sparse linear solvers. Fractional SPDE solvers or rational approximations can be developed, but they are not as readily available for this four-dimensional, inhomogeneous, anisotropic application. Applications requiring rougher fluctuations can replace the factorized SPDE with a fractional approximation, a different colored-noise forcing, or a calibration of the correlation tensor and fluctuation amplitude against GRMHD data.

Several extensions are immediate. The angular-momentum profile can be replaced by a different analytic or simulation-calibrated thick-disk profile, provided the normalization and timelike-domain checks are retained. Because the model parameters encode effective source properties (e.g., rotation and inflow) and the correlation scales, directions, and amplitudes set the field’s spatiotemporal structure, they could be calibrated to fluctuation statistics from GRMHD simulations (e.g., density, magnetic-field strength, emissivity). Alternatively, calibration could use observables by matching synthetic image sequences, such as pattern speeds, ring brightness structure, and variability. These comparisons could constrain effective parameters for a given radiative-transfer prescription and viewing geometry but may not uniquely map to underlying plasma properties.

Similarly, the jet velocity can be tied to a separate outflow prescription. The scalar emissivity can be generalized to polarized radiative coefficients, with correlated fluctuations in emissivity, absorptivity, and Faraday rotation. Finally, if the application requires statistically independent disk and jet variability, the same SPDE can be drawn twice with different tensors rather than using the default composite tensor. These extensions preserve the central structure of the model: a timelike Kerr emitter velocity coupled to a four-dimensional anisotropic Gaussian random field for time-dependent ray tracing (using codes such as~\texttt{GYOTO}~\citep{Vincent:2011wz} or ~\texttt{jipole}~\citep{Motta:2025nsv}). The values chosen for all the parameter in this work are merely illustrative and not tuned to any specific simulation or source. A systematic calibration would require broad simulation ensembles, synthetic observations, and a study of parameter degeneracies, which is beyond this work and is deferred to a separate study.

\begin{acknowledgments}
A.~C.-A. thanks T.~Gravely, H.~Lim, A.~Lupsasca, J.~Miller, and in particular R.~Falgout and C.~Gammie, for insightful discussions and comments on the numerical implementation developed in this work. We also thank the anonymous referee for their feedback, which enabled us to improve our results and manuscript presentation. DRG is supported by the Agencia Estatal de Investigación Grant Nos. PID2022-138607NB-I00 and CNS2024-154444, funded by MICIU/AEI/10.13039/501100011033 (Spain). Computations were performed using the Wake Forest University (WFU) High Performance Computing Facility, a centrally managed computational resource available to WFU researchers, including faculty, staff, students, and collaborators~\cite{WakeHPC}. 
\end{acknowledgments}

\bibliography{refs}{}
\bibliographystyle{aasjournal}

\end{document}